\documentclass[fleqn,twoside]{article}
\usepackage{espcrc2}
\usepackage{graphicx}
\usepackage[figuresright]{rotating}

\newcommand{\AmS}{{\protect\the\textfont2
  A\kern-.1667em\lower.5ex\hbox{M}\kern-.125emS}}

\newcommand{\les}{\stackrel{<}{{}_{\sim}}}

\newcommand{\VEV}[3]{\left\langle #1\left| #2 \right| #3\right\rangle}

\newcommand{\nn}{\nonumber}

\title{Calcutation of kaon matrix elements in quenched domain-wall QCD 
with DBW2 gauge action}

\author{J.~Noaki\address[RBRC]{RIKEN BNL Research Center, Brookhaven
National Laboratory, Upton, NY 11973-5000, USA}\ \  for RBC
Collabolation\thanks{
We thank RIKEN, BNL and the U.S.\ DOE for providing the facilities
essential for the completion of this work.}}

\begin{document}
\begin{abstract}
We give a progress report of our new $a^{-1}\approx 3$ GeV quenched 
calculation of kaon matrix elements with domain-wall fermion and DBW2 
gauge action. Our smaller lattice spacing allows us to address the 
effect of charmed quark on the lattice. We show preliminary results of 
$B_K$ renormalized non-perturbatively and  $K\to\pi$ matrix elements.
\vspace{1pc}
\end{abstract}
\maketitle

\section{Introduction}

In order to treat the QCD effect in non-leptonic kaon decays 
non-perturbatively, it is important for the matrix elements of the 
local operators to be calculated on the lattice. A couple of years ago,
CP-PACS and RBC Collaboration~\cite{CPPACSep,RBCep} calculated all 
of the matrix elements for the interaction of $K\to\pi\pi$ decay: 
$H_W=\sum_{i}C_i(\mu)Q_i$. Using the domain-wall fermion formalism to 
realize the chiral 
symmetry required in this calculation, they reported small and negative 
values of $\epsilon'/\epsilon$ in conflict with the experimental result. 
In these works, however, there are several uncontrolled systematic
errors coming from 1) the small, but non-zero breaking of chiral symmetry, 
2) finite lattice spacing, 3) the perturbative treatment of the charmed 
quark in the matrix elements, 
4) quenching effect, and 5) $K \to \pi\pi$ matrix elements obtained 
from $K \to \pi$ and $K \to 0$ by using lowest order chiral perturbation 
theory~\cite{BernardSoni}.

In order to examine the first three of above systematic errors extensively, 
we are performing a quenched simulation with domain-wall fermion and the 
DBW2 gauge action which improves the chiral symmetry on the 
lattice~\cite{RBCDBW2}. 
The degree of chiral symmetry breaking is decreased by a factor 1/10 
compared with the previous work of RBC Collaboration. 
In addition, the effect of the charmed quark on the lattice can be examined 
as well as the scaling violation.

In the rest of this article, we present the contents of the numerical 
simulation and report preliminary results of kaon B-parameter $B_K$ and 
the matrix elements which numerically dominate $\Delta I = 1/2$ rule 
and $\epsilon'/\epsilon$.
Our dynamical simulation to study the quenching effect 
is reported in refs. \cite{DYNAMICAL}.

\section{Numerical Simulation}
\begin{table}[b]
\caption{Simulation parameters and preliminary results of basic 
quantities.}
\label{PARAMS}
\begin{tabular}{rl}
\hline
size: & $24^3\times 48$\\
DBW2: & $\beta = 1.22$, \\
\#sweeps& 5,000 ($B_K$), 10,000 ($K\to\pi$, meson)\\
 DWF: & $M_5 = 1.65\ ,\ L_s = 10$ \\
$m_fa$ & 0.008 -- 0.040, in step of 0.008 \\
$m_ca$ & 0.08, 0.12, 0.16, 0.20, 0.30, 0.40, 0.50\\
\#configs. & 77 ($B_K$), 23 (NPR), 42 ($m_{\rm res}$), \\ 
&   50 ($K\to\pi$, meson)\\
\hline
$a^{-1}$& 2.86(9) GeV ($m_fa=0$, from $\rho$-meson)\\
$m_{\rm res}a$& $9.73(4)\cdot 10^{-5}$ ($m_fa=0$)\\
\hline
\end{tabular}
\end{table}
In Table.~\ref{PARAMS}, simulation parameters used in this calculation 
and preliminary results of basic quantities are summarized. 
Our strategy of gauge generation with a rather fine lattice spacing
and well-distributed topological charge is discussed in ref.~\cite{LAT02NOAKI}.
Since quark mass $m_fa$ is introduced as a parameter 
of the boundary condition in the fifth dimension in domain-wall QCD,
localization of chiral modes on both domain-walls in the fifth 
dimension tends to fail for a heavy quark mass.
However, our small lattice spacing made the value of $m_ca$ acceptable 
as a domain-wall fermion: $m_ca\simeq$ 0.45. We observed that, around this
value, the shape of the wave function in the fifth dimension was
qualitatively similar to that with much smaller $m_f$.
The small value of residual quark mass $m_{\rm res}\les$ 0.3 MeV 
demonstrates the good chiral symmetry. Therefore, it is expected that 
at least for those operators that do not mix with lower dimensional
ones, finite $L_s$ causes negligible chiral symmetry breaking effects.

\section{Kaon B-parameter $B_K$}

\begin{figure}[t]
\begin{minipage}{0.7\linewidth}
\includegraphics[width=4.9cm,clip]{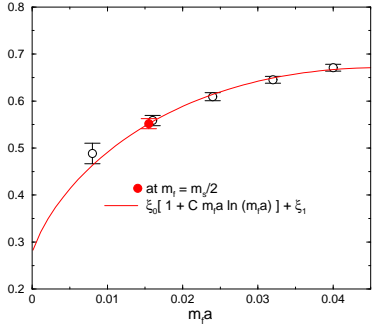}
\vspace*{-0.8cm}

 \caption{Lattice value of $B_K$ as a function of $m_fa$.}
\label{BK}
\end{minipage}
\end{figure}

$B_K$ on the lattice can be obtained as a ratio of the 
matrix elements of $Q_{\Delta S=2}= [\bar{s}\gamma_\mu(1-\gamma_5)d]^2$ and 
axial current $A_4$.
Our result is shown in Fig.~\ref{BK} as a function of $m_fa$.
The fit function used is 
$B_K= \xi_0[1+Cm_fa \ln (m_fa)] +\xi_1 m_fa$ with the coefficient $C$
taken from analytic result~\cite{Sharpe}.  In this chiral fit, 
$\chi^2/{\rm dof} = 0.76$ is obtained.

Taking only statistical errors into account, our preliminary results are 
summarized as
\begin{eqnarray}
& &B_K^{\rm latt}(m_fa=0.0155) = 0.552(11), \nn\\
& &B_K^{\rm RI\mbox{-}MOM}(\mu=2\ {\rm GeV}) = 0.542(11),\nn\\
& &B_K^{\rm \overline{\rm MS}, NDR}(\mu=2\ {\rm GeV}) = 0.549(11),\nn\\
& &\hat{B}_K= 0.764(15).\nn
\end{eqnarray}
The lattice value $B_K^{\rm latt}$ is quoted at $m_f = m_s/2$ 
(filled symbol in Fig.~\ref{BK}).
By a non-perturbative renormalization procedure following~\cite{NPR}, 
we obtained renormalization factor 
 $Z_{B_K}^{\rm RI\mbox{-}MOM}(\mu=2\ {\rm GeV}) = 0.9816(79)$.
In this step, we also estimated the degree of the mixing of the 
operators with wrong chirality. As expected by the small value of 
$m_{\rm res}$, all elements of the amputated four-point function in 
the chirality basis were observed to be less than 0.2\% of the 
$(VV+AA, VV+AA)$ element which corresponds to $Z_{B_K}$.
Though we shifted $B_K^{\rm RI\mbox{-}MOM}$ to 
$B_K^{\overline{\rm MS}}$~\cite{RItoMSb} and the renormalization group
invariant (RGI) value $\hat{B}_K$~\cite{Ciuchini} with $N_f=0$,
results will change less than 1\% even with $N_f=3$.
Our result of $B_K^{\overline{\rm MS}}(\mu=2\ {\rm GeV})$ is consistent 
with CP-PACS~\cite{CPPACSBK} ($a^{-1}\simeq$ 2.9 GeV, $24^3\times60\times16$)
and the previous work of RBC~\cite{RBCep}.

\section{$K\to\pi$ matrix elements}
\begin{figure}[t]
\hspace*{-2.7cm}
\begin{minipage}{1.35\linewidth}
\includegraphics[width=5.2cm,clip]{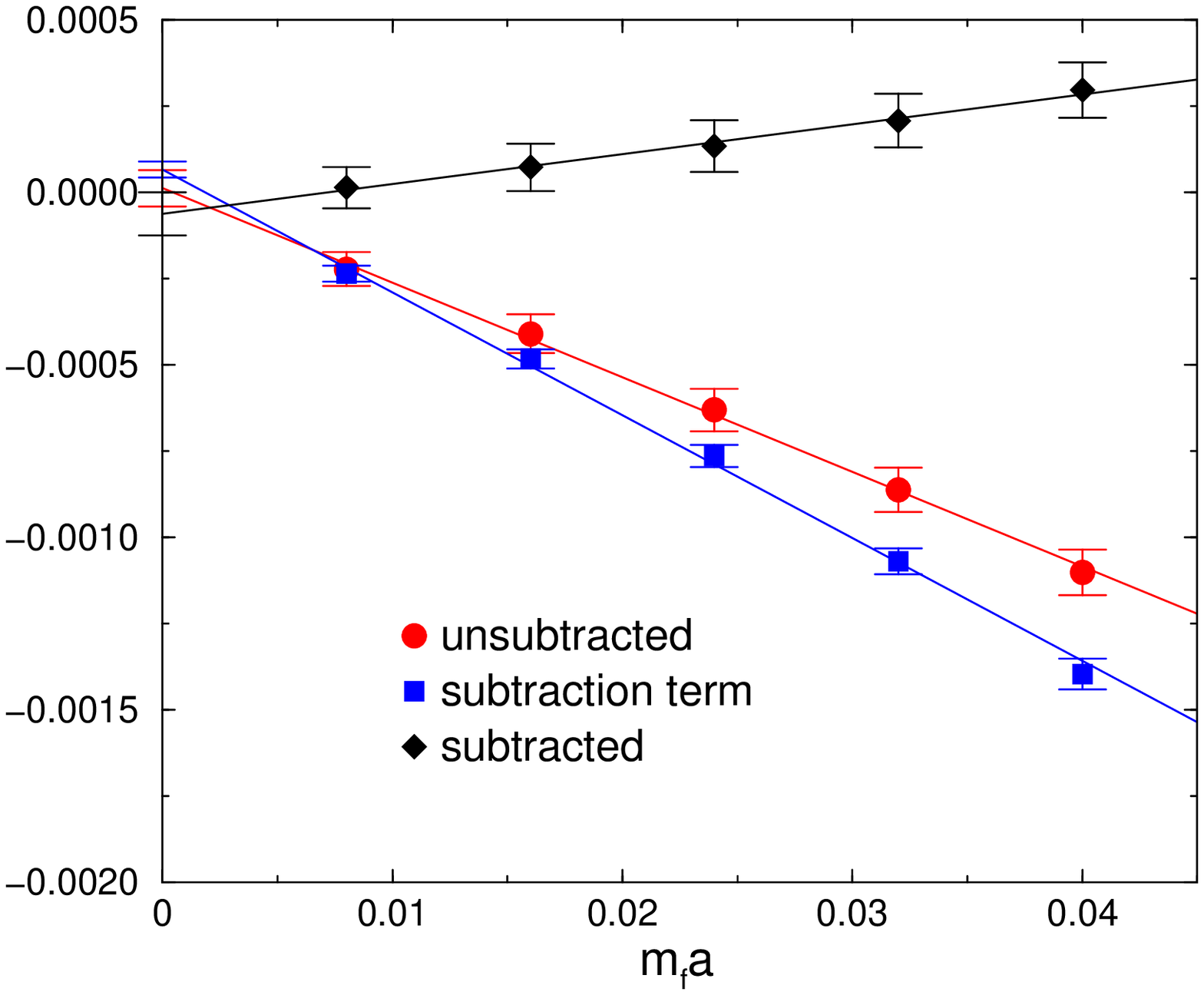}
\hspace{0.2cm}
\includegraphics[width=4.5cm,clip]{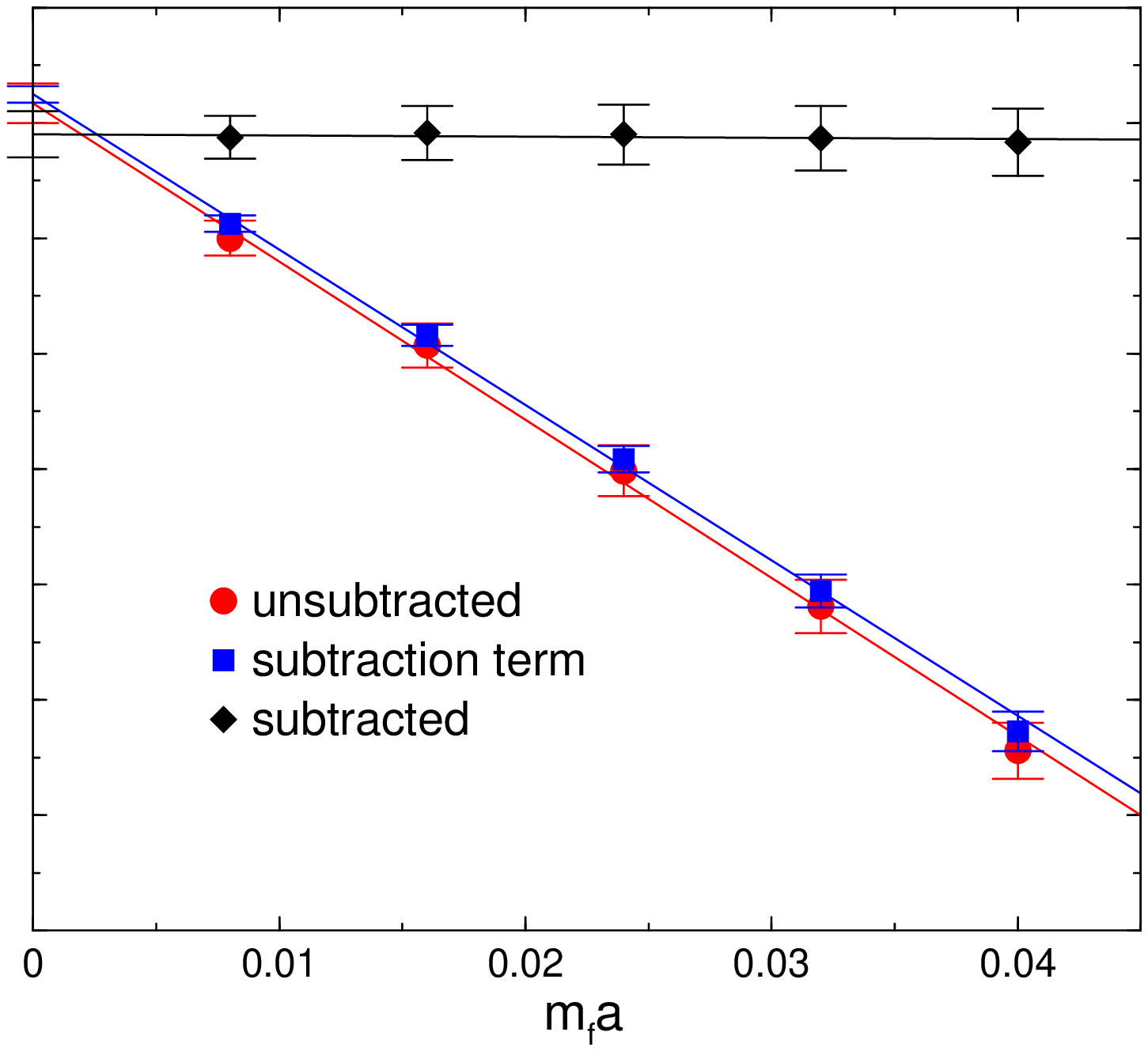}
\vspace*{-1.2cm}

\caption{$K\to\pi$ matrix elements of $Q^{(0)}_2$ (left) and
 $Q_{2c}$ (right, with $m_ca=0.40$) as a function of $m_fa$. 
Linear fit was used.}
\label{Q02}
\end{minipage}
\end{figure}
At the lowest order of chiral perturbation theory, $K\to\pi\pi$ matrix 
elements are proportional to $K\to\pi$ matrix elements calculated on
the lattice~\cite{BernardSoni}. For $i=1$ -- $6, 9$ and $10$, these 
matrix elements are related as
\begin{eqnarray}
& &\hspace{-0.5cm}\VEV{\pi^+\pi^-}{Q_i^{(I)}}{K^0}=\nonumber\\
& &\hspace{-0.7cm}\frac{m_K^2-m_\pi^2}{\sqrt{2}f}
\left[\frac{\VEV{\pi^+}{Q_i^{(I)}}{K^+}_{\rm sub}}{m_{\rm PS}^2}
+{\cal O}(m_{\rm PS}^2)\right],
\end{eqnarray}
where $Q_i^{(I=0,2)}$ is the contribution to the final state with $I=0,2$.
Only for the case of $\Delta I = 1/2$, or $I =0$, subtraction
is needed to resolve mixing with a lower dimensional operator:
\begin{eqnarray}
\VEV{\pi^+}{Q_i^{(0)}}{K^+}_{\rm sub}\!\!\!\!=\!
\VEV{\pi^+}{Q_i^{(0)}-\alpha_i Q_{\rm sub}}{K^+},\label{subtraction}
\end{eqnarray}
\begin{eqnarray}
Q_{\rm sub}&\equiv& (m_s+m_d)\bar{s}d-(m_s-m_d)\bar{s}\gamma_5d,\\
\alpha_i&=& \VEV{0}{Q_i^{(0)}}{K^0}/\VEV{0}{Q_{\rm sub}}{K^0}.
\end{eqnarray}
In Fig.~\ref{Q02}, $K\to\pi$ matrix elements of $Q^{(0)}_2$ (left) and 
$Q_{2c}=(\bar{s}u)_L(\bar{u}d)_L$ (right) are 
presented before and after the subtraction of (\ref{subtraction}).
In particular, one finds that the slope of the subtracted 
matrix element of $Q_{2c}$ is much smaller than that of $Q^{(0)}_2$, 
which might mean a minor contribution of $\VEV{\pi\pi}{Q_{2c}}{K}$ to 
$\Delta I=1/2$ rule. 

\begin{figure}[t]
\begin{minipage}{1.3\linewidth}
\includegraphics[width=5.0cm,clip]{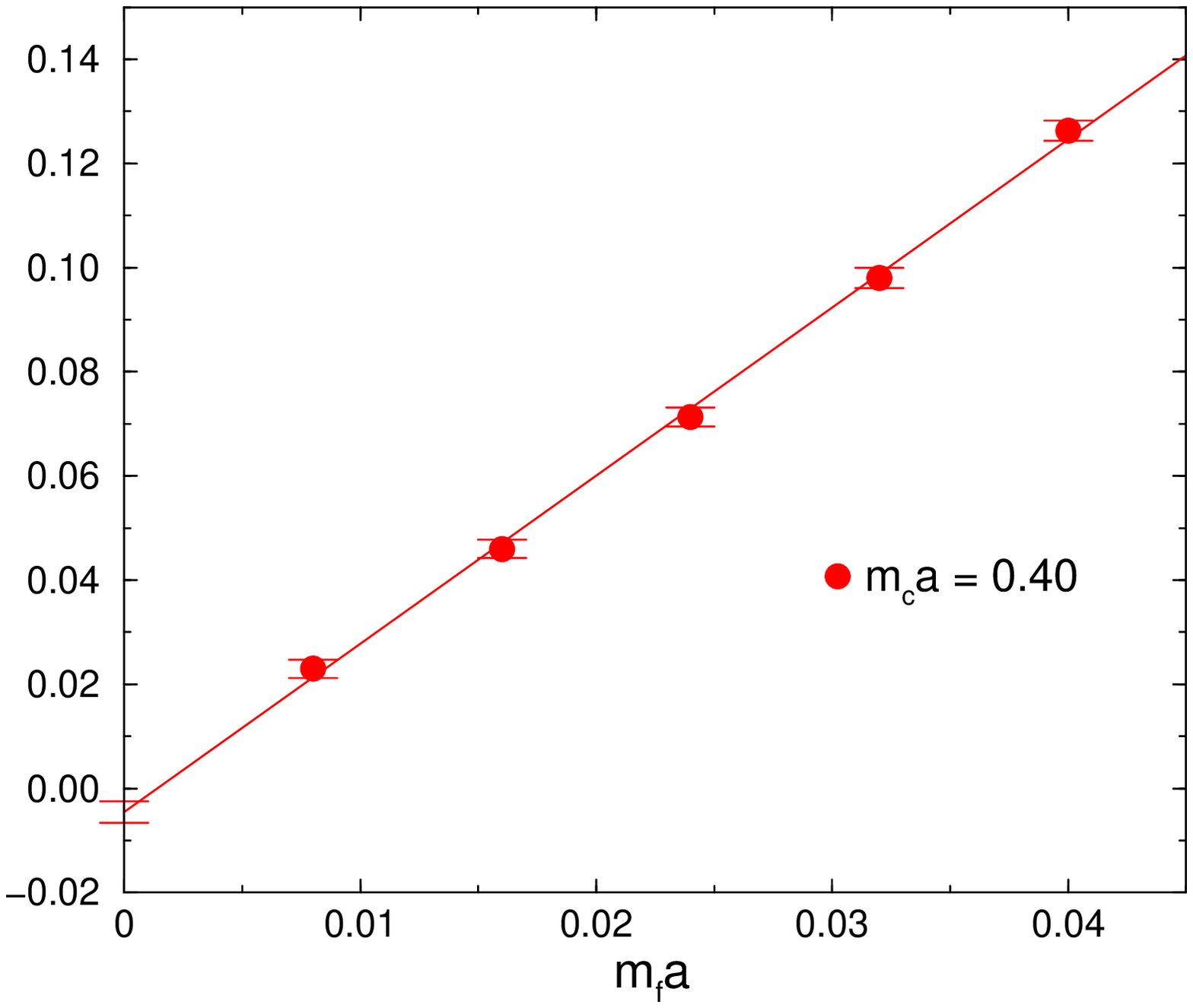}
\includegraphics[width=5.1cm,clip]{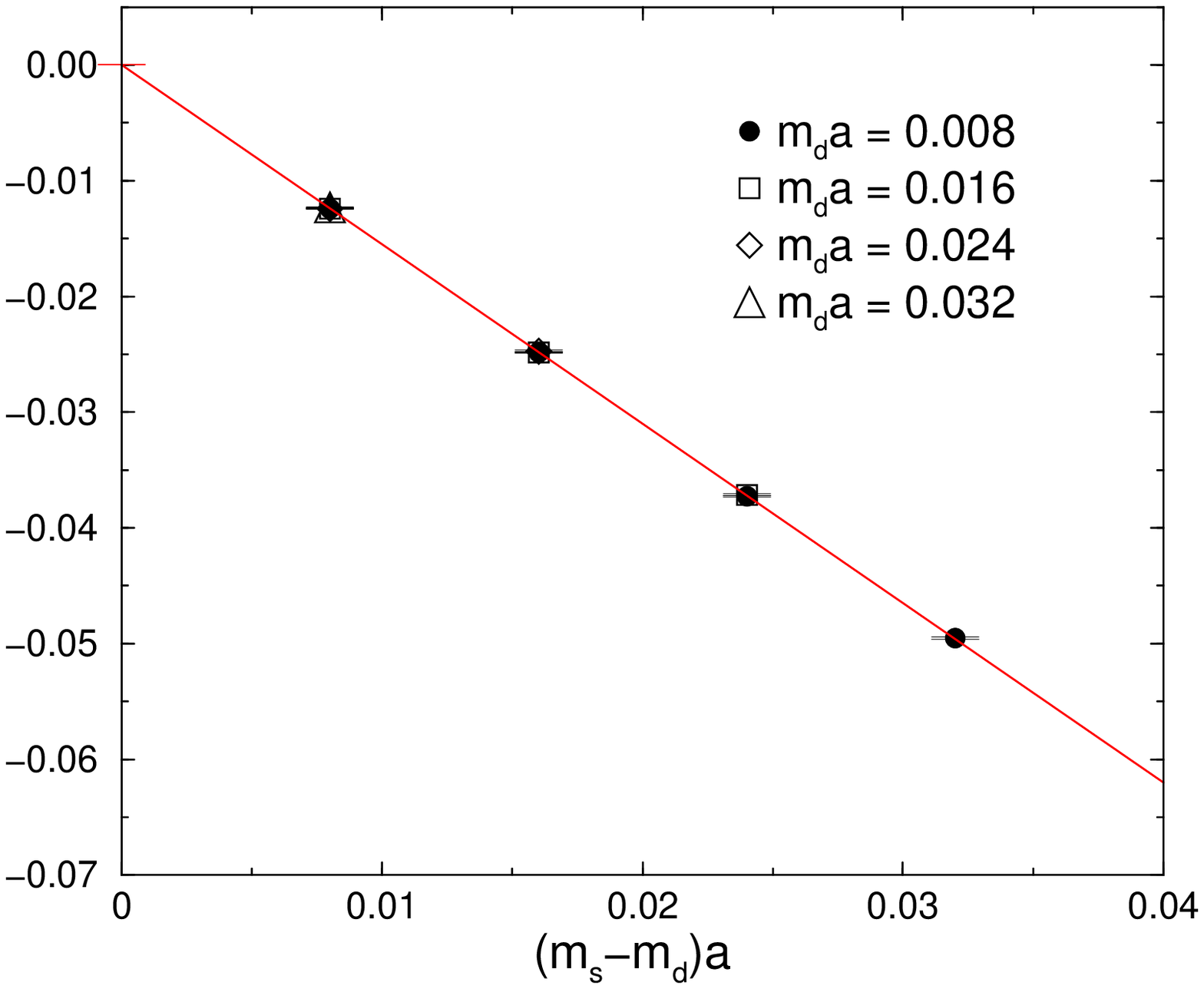}
\vspace{-1.2cm}

\caption{Unsubtracted $K\to\pi$ matrix element of $Q^{(0)}_6$ and 
a ratio of $K\to0$ matrix elements of $Q^{(0)}_6$ and $\bar{s}\gamma_5d$. 
Linear fit was used for both cases. }
\label{Q06}
\end{minipage}
\vspace{-0.1cm}

\end{figure}
Fig.~\ref{Q06} shows $\VEV{\pi^+}{Q^{(0)}_6}{K^+}$ without the subtraction 
(left) and $\VEV{0}{Q^{(0)}_6}{K^0}/\VEV{0}{\bar{s}\gamma_5 d}{K^0}$ as a 
function of $m_sa-m_da$ (right) whose slope should be $\alpha_6$.
Though these are an example with $m_ca=0.40$, their dependence on 
$m_ca$ is not visible, so far. 
While the slopes of both quantities are determined within the error of 
2\% and 0.2\% respectively, $K\to\pi\pi$ matrix element which is obtained as
a combination of them has the error more than $200\%$ due to a severe 
subtraction. Therefore, before we can quote result for
$\VEV{\pi^+\pi^-}{Q^{(0)}_6}{K^0}$, we need to significantly improve our
statistics. 
\begin{figure}[t]
\hspace{2.0cm}
\begin{minipage}{0.7\linewidth}
  \includegraphics[width=5.0cm]{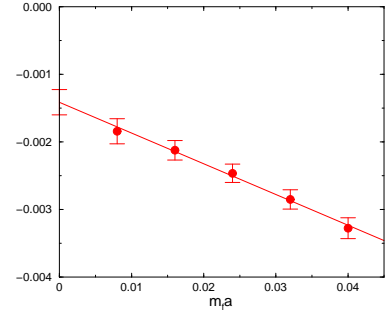}
\vspace{-0.9cm}

  \caption{$K\to\pi$ matrix element of $Q^{(2)}_8$ as a function of
 $m_fa$ with its liniar fit.}
\label{Q28}
\end{minipage}
\end{figure}

For $i=7$ and $8$, $K\to\pi\pi$ and $K\to\pi$ matrix elements are
related as
\begin{eqnarray}
\VEV{\pi^+\pi^-}{Q_i^{(I)}}{K^0}&=&-\frac{1}{\sqrt{2}f}
\VEV{\pi^+}{Q_i^{(I)}}{K^+}\nonumber\\
& &+{\cal O}(m_{\rm PS}^2).
\end{eqnarray}
We show $\VEV{\pi^+}{Q_8^{(2)}}{K^+}$ in Fig.~\ref{Q28} and observe its 
expected shape and intercept roughly consistent with previous 
works~\cite{CPPACSep,RBCep}.
\newcommand{\NP}{Nucl.~Phys.}
\newcommand{\NPSup}{Nucl.~Phys.~{\bf B} (Proc.~Suppl.)}
\newcommand{\PL}{Phys.~Lett.}
\newcommand{\PR}{Phys.~Rev.}
\newcommand{\PRL}{Phys.~Rev.~Lett.}


\begin{thebibliography}{15}
\bibitem{CPPACSep} J.~Noaki {\it et al.} (CP-PACS Collaboration), 
                    Phys. Rev. D{\bf 68} (2003) 014501.
\bibitem{RBCep} T.~Blum {\it et al.} (RBC Collaboration),
	hep-lat/0110075, to appear in Phys. Rev. D.
\bibitem{BernardSoni} C.~Bernard {\it et al.}, Phys. Rev.{\bf D32}(1985) 2343.
\bibitem{RBCDBW2} Y.~Aoki {\it et al.} (RBC Collaboration), hep-lat/0211023.
\bibitem{DYNAMICAL} See contributions of C.~Dawson, T.~Izubuchi and 
R.~Mawhinney to these proceedings.
\bibitem{LAT02NOAKI}J. Noaki for RBC Collaboration, Nucl.~Phys. {\bf B} 
(Proc.~Suppl.) 119 (2003) 362.
\bibitem{Sharpe} Sharpe, S., Phys. Rev. {\bf D46} (1992) 3146.
\bibitem{CPPACSBK} A.~Ali Khan {\it et al.}
(CP-PACS Collaboration), Phys. Rev. {\bf D64} (2001) 114506.
\bibitem{NPR}G.~Martinelli {\it et al.}, Nucl. Phys. {\bf B445} (1995)	81;
T.~Blum {\it et al.} (RBC Collaboration), Rhys. Rev. D {\bf 66} (2002) 014504.
\bibitem{RItoMSb} M.~Ciuchini {\it et al.}, Z. Phys. {\bf C68} (1995) 239.
\bibitem{Ciuchini} M.~Ciuchini {\it et al.}, Nucl. Phys. {\bf B523} (1998) 501.
\end{thebibliography}
\end{document}